\documentclass[aps,prb,noshowpacs,footinbib,superscriptaddress,twocolumn,floatfix]{revtex4-1}

\usepackage{amsbsy}
\usepackage{amsmath}
\usepackage{amssymb}
\usepackage{amsfonts}
\usepackage{graphicx}
\usepackage{pst-all}
\usepackage[english]{babel}
\usepackage{color}
\usepackage[normalem]{ulem}
\usepackage{bm}
\usepackage{yfonts}

\begin{document}

\date{\today}

\title{On the Interpretation of Thermal Conductance of the $\nu=5/2$ Edge}

\author{Steven H. Simon}
\affiliation{Rudolf Peierls Centre for Theoretical Physics, 1 Keble Road, Oxford, OX1 3NP, UK}

\begin{abstract}

Recent experiments [Banerjee et al, arXiv:1710.00492] have measured thermal conductance of the $\nu=5/2$ edge in a GaAs electron gas and found it to be quantized as $K \approx 5/2$ (in appropriate dimensionless units).  This result is unexpected, as prior numerical work predicts that the $\nu=5/2$ state should be the Anti-Pfaffian phase of matter, which should have quantized $K=3/2$.  The purpose of this paper is to propose a possible solution to this conflict: if the Majorana edge mode of the Anti-Pfaffian does not thermally equilibrate with the other edge modes, then $K=5/2$ is expected.    I briefly discuss a possible reason for this nonequilibration, and what should be examined further to determine if this is the case.  
\end{abstract}

\pacs{PACS}

\maketitle


Perhaps the most well-known feature of quantized Hall states is their quantized Hall conductance.   If we think in the language of edge-state transport we can define the conductance of an edge $G = e  \partial J/\partial \mu$ as the change in current along the edge when the chemical potential is raised.   When a state is well quantized, meaning it is gapped in the bulk, the conductance $G$ is quantized as $G= \nu e^2/h$ where $\nu$ is the filling fraction, an integer or simple fraction.   

We can similarly define the thermal conductance along an edge as $G_Q = \partial J_Q/\partial T$ where $J_Q$ is the thermal current running along the edge, and $T$ is the temperature of the edge.      A classic theoretical work by Kane and Fisher\cite{KaneFisherThermal} predicted that the thermal conductance of a quantum Hall edge should also be quantized.   If we write 
$
G_Q = K \kappa_0 T
$
with $\kappa_0 = \pi^2 k_B^2/(3 h)$,  the dimensionless thermal conductance coefficient $K$, under certain conditions, should be given by the central charge of the edge (an integer or simple fraction)\cite{KaneFisherThermal,Capelli}. For Abelian quantum Hall states, all of the edge modes are bosonic with unit central charge.  Each mode therefore contributes one unit to the coefficient $K$.   For non-Abelian states, the central charge of an edge mode can be a simple fraction different from one\cite{Capelli}.  In particular a Majorana edge mode contributes one half unit to $K$.    In cases where multiple edge modes all propagate in the same direction, the contributions of the different edge modes simply add.   In cases where edge modes propagate in both directions,   if the edges thermally equilibrate, then $K$ is the  difference in the central charges of the downstream (right moving) modes minus the upstream (left moving) modes\cite{KaneFisherThermal}.

While it may have once seemed technologically impossible to measure the thermal conductance of a quantum Hall state, recent experimental advances have shown that it can be done\cite{Jezouin601,BanerjeeAbelian,BanerjeeNonAbelian}.   A spectacular set of experiments\cite{BanerjeeAbelian,BanerjeeNonAbelian} recently confirmed predictions of the thermal conductance for several Abelian quantum Hall states, as shown in Table \ref{tab:Abelian}.   The good agreement between theory and experiment is strong validation that the experimental method is accurately measuring thermal conductance of the quantum Hall state and not other effects such as phonons. 

\begin{table}
\vspace*{10pt}
\hspace*{.2cm}
{
\begin{tabular}{c|c|c}
$\nu$  & predicted $K$ & measured $|K|$   \\
\hline 
 2 &   $\phantom{-}2 = 2 - 0$  & 1.96 $\pm$   .06
 \\
 1 &  $\phantom{-}1 = 1 - 0$  & .90 $\pm$   .09  \\
 1/3 &   $\phantom{-}1 = 1 - 0$  & 1.00 $\pm$   .045  \\
 2/3 & $\phantom{-} 0 = 1 - 1$ & 0.328 $\pm$ .024  See text  \\
 3/5 &   $-1 = 1 - 2$ & 1.040 $\pm$   .041  \\
 4/7 &   $-2 = 1 - 3$ & 2.045 $\pm$   .052  \\
7/3 &  $\phantom{-}3 = 3 - 0$ & 2.86  $\pm$   .03  \\
8/3 &  $\phantom{-}2 = 3 - 1$ & 2.11 $\pm$   .01     
\end{tabular}
\\ 
}
\caption{Dimensionless thermal conductances for Abelian quantum Hall states. For Abelian states, all edge modes contribute one unit to the dimensionless thermal  conductance $K$.  Here  $K$ is written as number of downstream modes minus the number of upstream modes.  Note that the experiment measures $|K|$ rather than $K$.  Experimental results are quoted from Ref.~\onlinecite{BanerjeeAbelian} except for $\nu=7/3$ and $8/3$ which are from Ref.~\onlinecite{BanerjeeNonAbelian}.  }
\label{tab:Abelian}
\end{table}

Note that the measured thermal conductance of the $\nu=2/3$ state appears to be a bit in disagreement with prediction.   One possible explanation given by the authors of Ref.~\onlinecite{BanerjeeAbelian} is that the edge modes are not fully thermally equilibrated, despite the length of the edge being fairly long (roughly 150 microns).     At slightly higher temperature, where equilibration should be more effective, the measured value of $|K|$ drops to $\approx .25$, potentially suppporting this explanation.    Non-equilibration may also explain\cite{BanerjeeNonAbelian} the small error in $\nu=8/3$. 

To understand the effect of non-equilibration, it is worth considering a limit where the edge modes do not equilibrate at all.  In this case, the thermal conductances of the edge modes add in absolute value (like resistors in parallel).   For $\nu=2/3$ this would predict a measured $|K|=2$ and for $\nu=8/3$ this would predict a measured $|K|=4$.   Thus, lack of full equilibration could raise both of the measured values slightly from their predicted fully equilibrated values, as observed in the experiment. 

We now turn to the main subject of this paper: thermal conductance of the $\nu=5/2$ state.  The crucial experimental observation\cite{BanerjeeNonAbelian} is that the thermal conductance is quantized at $K \approx 5/2$, the measured conductance varying from $|K|=2.55 \pm .01$ at 18mK to roughly $2.76$ at 10 mK.   The authors of Ref.~\onlinecite{BanerjeeNonAbelian} argue that $|K|$ is well quantized at $5/2$ at the higher temperature and shifts slightly at lower temperature due to imperfect thermal equilibration.

\begin{table}
\vspace*{10pt}
\hspace*{.2cm}
{
\begin{tabular}{c|c|c|c|c|c}
possible states  ~~~  & 
\rotatebox{90}{$\!\!\!\!\!\!\!\!\!$downstream}$\,$ \rotatebox{90}{$\!\!\!\!\!\!\!\!\!$Bose}& \rotatebox{90}{$\!\!\!\!\!\!\!\!\!$upstream}$\,$\rotatebox{90}{$\!\!\!\!\!\!\!\!\!$Bose} & \rotatebox{90}{$\!\!\!\!\!\!\!\!\!$downstream}$\,$\rotatebox{90}{$\!\!\!\!\!\!\!\!\!$Majorana} & \rotatebox{90}{$\!\!\!\!\!\!\!\!\!$upstream}$\,$\rotatebox{90}{$\!\!\!\!\!\!\!\!\!$Majorana} & predicted $K$ 
\\ 
of $\nu=5/2$ & &  &  &  &   ({\tiny thermally equilibrated})\\
 \hline
 $SU(2)_2$ & 4 & 0 & 1 & 0 & 9/2 \\
Moore-Read & 3 & 0 & 1 & 0 & 7/2   \\ 
PH-Pfaffian & 2 & 0 & 1 & 0 & 5/2   \\ 
Anti-Pfaffian & 3 & 1 & 0 & 1  & 3/2  \\
Anti-$SU(2)_2$  & 3 & 2 & 0 & 1  & 1/2 
\end{tabular}}
\caption{Proposed states for $\nu=5/2$, and their predicted dimensionless thermal conductances assuming full thermal equilibration of all of the edge modes.   In Ref.~\onlinecite{BanerjeeNonAbelian}, the measured thermal conductance at $\nu=5/2$ ranges from $|K|=2.55 \pm .01$ at 18mK to roughly $2.76$ at 10 mK.    Bose modes contribute 1 unit to $K$ and Majorana modes contribute 1/2.   Upstream modes contribute with a negative sign assuming full thermal equilibration (but contribute with a positive sign if they are not thermally equilibrated).  We list here only states that include Majorana modes (states such as the  331 state are not listed as they predict integer $K$).  
 }
\label{tab:nonabelian}
\end{table}

A clear measurement of $|K|$ being quantized at a number which is not an integer would be strong evidence of a non-Abelian quantum Hall state.   The leading proposals for the nature of $\nu=5/2$ all predict that $K$ should be a half-integer due to a non-Abelian Majorana edge mode which has a central charge of 1/2.   Thus the observation by Ref.~\onlinecite{BanerjeeNonAbelian} of this half-quantum of thermal conductance is an extremeley exciting result.    However, there remains a major puzzle: the integer part of $K$ is unexpected (See the discussion in Ref.~\onlinecite{HalperinJournalClub}).     

In Table \ref{tab:nonabelian} we show the candidate nonabelian states that are considered by Ref.~\onlinecite{BanerjeeNonAbelian}.  The authors argue that the most likely state of matter to describe $\nu=5/2$ is the so-called PH-Pfaffian\cite{Son,Fidkowski} --- a state of matter with particle-hole symmetry within the Landau level.  Although the measured value of $|K|\approx 2.5$ does appear to match the prediction for the PH-Pfaffian, there are some serious problems with drawing this conclusion which we now discuss. 

For almost 20 years, since the first substantial numerical exact diagonalization studies\cite{Morf} of the $\nu=5/2$ state, evidence has been very strong that either the Moore-Read state\cite{Moore} or Anti-Pfaffian state \cite{Levin,LeeAntiPfaffian} is the incompressible ground state of $\nu=5/2$ in high mobility GaAs heterostructures\cite{Morf,Storni,Wang,RezayiHaldane}.   These two possibilities are particle-hole conjugates of each other, having the same energy if particle-hole conjugation remains a good symmetry.   Landau level mixing (not considered in early numerical work) is required to break the degeneracy between these two possibilities.   Recent numerical work has clarified that the Anti-Pfaffian should be favored in the physical systems once Landau level mixing is accounted for properly\cite{RezayiRecent,RezayiSimon,Zaletel}.   In no numerical work has the PH-Pfaffian ever appeared as a ground state of a disorder-free system.  Indeed, it does not appear as if it is remotely competitive. 

An interesting possibility suggested by Ref.~\onlinecite{Zucker} is that disorder might stabilize the PH-Pfaffian --- and this would have been missed in small system diagonalizations without disorder.   One may imagine that disorder breaks particle-hole symmetry locally and the system splits up into domains of Moore-Read state and domains of  Anti-Pfaffian, with internal edge states running along the domain boundaries.  The physics of such a network of internal edge states has been studied in detail in two recent works\cite{Mross,Ashvin}.  The outcome of the calculations depend  somewhat on the input parameters and assumptions, but there are two main outcomes that are common:   First, a first order transition between Pfaffian and Anti-Pfaffian is a possibility.  Second, a so-called thermal metal is possible --- a  phase with quantized electrical Hall conductance, but unquantized thermal conductance\cite{Mross,Ashvin,WanYang}.  Neither of these would match the result of experiment.   While both  recent works\cite{Mross,Ashvin} do find that, with sufficient assumptions, it is possible to stabilize a PH-Pfaffian-like state with quantized $K=5/2$, these additional assumptions do not appear realistic for the experiment (See in particular the discussion in Ref.~\onlinecite{Ashvin}).   Thus the experimental results remain a puzzle\cite{HalperinJournalClub}.  It has been suggested\cite{Ashvin,HalperinJournalClub} that the community might re-examine the numerical work in the absence of disorder which has previously predicted that the Anti-Pfaffian should be the ground state, to see if the PH-Pfaffian might actually be better for some range of parameters close to that of the  experiment.     While re-confirmation of previous work may be useful, considering the substantial effort that has been devoted to numerically examining $\nu=5/2$ already\cite{Morf,Storni,RezayiSimon,Zaletel,RezayiHaldane,RezayiRecent,Wang} it seems quite unlikely that such a major oversight has occurred.  

The purpose of this paper is to briefly point out that there is potentially a fairly simple interpretation that could bring the experimental observation\cite{BanerjeeNonAbelian} of $K\approx 5/2$ into agreement with numerical work predicting an Anti-Pfaffian ground state\cite{Morf,Storni,RezayiSimon,Zaletel,RezayiHaldane,RezayiRecent,Wang}.  The Anti-Pfaffian has a single upstream Majorana edge mode (along with four bose modes, three runing downstream and one running upstream).   If we assume that the bose modes all thermally equilibrate with each other, but the Majorana mode does not thermally equilibrate, then we should predict $K=5/2$ (as observed in experiment) rather than $K=3/2$ as presented in Table \ref{tab:nonabelian}.   To remind the reader, when edge modes do thermally equilibrate, the signed central charges add, but when they do not thermally equilibrate, they add in absolute value.   Thus in the case of the Anti-Pfaffian if the bose modes all equilibrate they contribute $2=3-1$ to $K$.  The Majorana mode contributes 1/2 to $K$, but, this now comes with a positive sign if the mode is out of thermal equilibrium with the other modes. 

The fact that thermal non-equilibration can be an issue is already established by the experiments at $\nu=2/3$ and possibly $8/3$, as mentioned above.  However in those cases we might say that the edge mode is {\it mostly} thermally equilibrated since the resulting thermal conductances are fairly close to those predicted if thermal equilibration is complete.   Nonetheless, these two cases point out that even for an edge which runs over a fairly long length scale (150 microns) it is quite possible for equilibration to be poor.    Further, since the behavior of a Majorana edge mode could potentially be different from that of a bose edge mode, it could easily be imagined that the bose modes equilibrate well whereas the Majorana mode does not.   I will elaborate on a possible reasons for this below. 

Let us also note that not much is  known about the detailed structure of an Anti-Pfaffian edge in realistic conditions.    To my knowledge there have been no numerical works analyzing this situation.   However, there has been a serious numerical attempt to understand the detailed structure of the closely related Moore-Read edge in Ref.~\onlinecite{RezayiWanYang} which predicts that for experimentally realistic geometries, the Majorana mode velocity can be much less than that of the bose mode (a factor of 6-8 lower).   While similar calculations have not been done for the Anti-Pfaffian edge, we might reasonably guess that the Majorana edge velocity could be quite small compared to that of the other edge modes. 

Let us now think in a bit more depth about equilibration between edge modes.   In the absence of disorder, scattering processes between edge modes simply renormalize edge velocities\cite{KaneFisher,KaneFisherThermal,KaneFisherPolchinski,Levin,LeeAntiPfaffian}.
Thus equilibration between edge modes relies on disorder.    If we assume that the disorder is relatively long wavelength (as is often assumed for high mobility heterostructures) then the disorder cannot change the wavevector of an edge mode very much in each discrete scattering event.    If the velocity of an edge mode is very low, this means that many scattering events would be requried to transfer substantial energy into the edge mode, so we might expect that low velocity edge modes equilibrate badly compared to high velocity modes. 

It is important to note that if charge does not equilibrate between edge modes then the electrical conductance will generally be unquantized\cite{KaneFisher,KaneFisherPolchinski}.   Thus given that the experimental has well quantized electrical Hall conductance, we have to assume that charge is equilibrating, but heat is not.  The scattering process of interest\cite{LeeAntiPfaffian,Levin} is  a disorder mediated process that  removes an electron from one of the downstream bosonic modes and transfers it into a combination of the upstream modes --- the upstream moving electron being made of a combination of a bosonic charge mode and the neutral Majorana mode.    Assuming again that the wavevector provided by the disorder is small (due to the disorder being smooth), and assuming that the velocity of the Majorana mode is very low compared to that of the bosonic modes, such a scattering process could only transfer a small fraction of the initial energy into the Majorana mode --- the bose mode absorbing most of the energy.  Thus, one might expect a regime where the charge and heat are well equilibrated between the bose modes, but the Majorana mode remains out of thermal equilibrium. 

It is interesting to examine the temperature dependence of the experimental data.   As mentioned above, the thermal conductance $|K|$ deviates upwards from $5/2$ at the lowest temperatures.  This could be explained if one assumed that at these temperatures the upstream bose mode is also starting to thermally equilibrate imperfectly (if it does not equilibrate at all, the thermal conductance would be $K=9/2$).   One might imagine that at higher temperatures than the experiment reports, the Majorana mode might start to thermally equilibrate and $K$ would start to drop down towards its expected equilibrated value of $K=3/2$. 

Ideally, to theoretically establish the mechanism I propose, one would want a detailed numerical simulation of an Anti-Pfaffian edge, in realistic geometries, as mentioned above.   Further, one would want to evaluate disorder induced scattering matrix elements between the various edge modes.   Such simulations are probably beyond what is currently possible with exact diagonalization, although potentially DMRG techniques\cite{Zaletel} may be helpful. 

In the absence of more detailed numerical simulations, it might be difficult for theory to be definitive about whether the mechanism proposed here applies  to the experiments of Ref.~\onlinecite{BanerjeeNonAbelian} (although it may be the only proposal that currently explains the experiments with no obvious contradictions).   However, there is a very straightforward experimental approach, suggested to me by Bertrand Halperin\cite{HalperinDiscussion}, which could be used.  If one were
to  add thermally and electrically floating contacts along the edges, these should tend to enhance equilibration of all of
the edge modes. The result of such an experiment should
be that the measured thermal conductance should move
towards $|K|  = 3/2$ appropriate for the thermally equilibrated Anti-Pfaffian.

In summary, I have proposed a mechanism by which the Anti-Pfaffian state of matter can display thermal conductance $K=5/2$ in agreement with experimental observation.   This mechanism assumes that the Majorana edge mode remains out of thermal equilibrium with the bosonic edge modes. 

Acknowledgements: The author would like to thank Bertrand Halperin for helpful conversations.  This work was supported by EPSRC grants
EP/I031014/1 and EP/N01930X/1. Statement of compliance with EPSRC policy framework on research data:
This publication is theoretical work that does not require
supporting research data.

\bibliographystyle{apsrev4-1}
\bibliography{Thermal}

\end{document}